\documentclass[twocolumn,prb,aps,showpacs,superscriptaddress]{revtex4}
\usepackage{graphicx}% Include figure files

\begin{document}

\title{Evidence for Induced Magnetization in Superconductor-Ferromagnet Hetero-structures: a Scanning Tunnelling Spectroscopy Study}

\author{Itay Asulin}
\affiliation{Racah Institute of Physics, The Hebrew University,
Jerusalem 91904, Israel}

\author{Ofer Yuli}
\affiliation{Racah Institute of Physics, The Hebrew University,
Jerusalem 91904, Israel}

\author{Gad Koren}
\affiliation{Department of Physics, Technion - Israel Institute of
Technology, Haifa 32000, Israel}

\author{Oded Millo}
\email{milode@vms.huji.ac.il} \affiliation{Racah Institute of
Physics, The Hebrew University, Jerusalem 91904, Israel}

\begin{abstract}
We performed scanning tunneling spectroscopy of c-axis oriented
$YBa_{2}Cu_{3}O_{7-\delta}$ films on top of which ferromagnetic
$SrRuO_{3}$ islands were grown epitaxially \textit{in-situ}. When
measured on the ferromagnetic islands, the density of states
exhibits small gap-like features consistent with the expected
short range penetration of the order parameter into the
ferromagnet. However, anomalous split-gap structures are measured
on the superconductor in the vicinity of ferromagnetic islands.
This observation may provide evidence for the recently predicted
induced magnetization in the superconductor side of a
superconductor/ ferromagnet junction. The length scale of the
effect inside the superconductor was found to be an order of
magnitude larger than the superconducting coherence length. This
is inconsistent with the theoretical prediction of a penetration
depth of only a few superconducting coherence lengths. We discuss
a possible origin for this discrepancy.
\end{abstract}


\maketitle
\section{Introduction}
Singlet pairing superconductivity and itinerant ferromagnetism are
two competing orders. The interplay between these orders has been
the focus of extensive experimental and theoretical works over the
past years. In the vicinity of a high quality interface between a
superconductor (S) and a ferromagnet (F) these two rival orders
are forced to coexist and thus allow the direct investigation of
their mutual influence. It is well established that in such
proximity systems, superconducting correlations can penetrate into
the ferromagnet and give rise to a rapid (nm scale) oscillatory
decay of the induced order parameter (OP) in F as a function of
the distance from the S/F interface. \cite{Buzdin review} A much
less studied aspect of the S-F proximity effect is the so called
"inverse proximity" effect. Here, fundamental questions arise: can
ferromagnetic order penetrate the S and how will such a
penetration affect superconductivity in the vicinity of the
interface? This subject has been the focal point of recent
theoretical works. \cite{Bergeret EPL66, Bergeret PRB70, Bergeret
PRB69, Bergeret PRB72, Krivoruchko, Kharitonov, Halterman, Cottet}
Although different approaches were used in these studies, all
share the basic conclusion that a sizable magnetic moment should
penetrate into the S side over the length scale of the
superconducting coherence length ($\xi_S$). The sign of this
moment, its spatial behavior and the actual mechanism underlying
this effect are still controversial. In general, according to the
quasiclassical approach that was employed in references,
\cite{Bergeret EPL66, Bergeret PRB70, Bergeret PRB69, Bergeret
PRB72, Krivoruchko, Kharitonov, Halterman} there are two possible
profiles for the induced magnetic moment inside the S. We will
refer to these as "screening" and "anti-screening" following the
terminology that was used in references \cite{Bergeret EPL66,
Bergeret PRB70}. In the screening profile, the direction of the
induced magnetic moment inside the S is anti-parallel to the
magnetization of F whereas in the anti-screening scenario it is
parallel. For a specific S/F bilayer, the sign of the induced
magnetic moment inside the S is influenced by several parameters:
the strength of the exchange field in F, the interface
transparency, the thickness of the F layer, the sign of the spin
polarization in F (the relative spin-polarization at the Fermi
surface with respect to the overall magnetization) and whether the
system is in the clean or diffusive limit. A change in one of
these parameters will lead to a change in the amplitude and even
the sign of the induced magnetic moment inside the S. In both
scenarios, the induced magnetic moment in S is predicted to decay
monotonically as a function of distance from the interface over
the length-scale of $\xi_{s}$. A different picture is presented in
Ref. \cite{Krivoruchko} in which the magnitude and sign of the
induced magnetization in S exhibits a damped oscillatory behavior
versus distance from the S/F interface on a scale of $\xi_{s}$.
According to Refs. \cite{Bergeret EPL66, Bergeret PRB70, Bergeret
PRB69, Bergeret PRB72, Halterman}, the induction of a magnetic
moment inside the S in such proximity systems is related to the
existence of an induced triplet pairing component at the S/F
interface. The possible formation of such a triplet pairing
component at the interface, its nature and its experimental
identification are one of the open issues in the field. A
different approach to the proximity effect in diffusive S/F
hybrids was employed by Cottet, where the spin dependence of the
phase shifts (SDIPS) acquired by electrons upon scattering on the
boundaries of the F layer are taken into account. \cite{Cottet}
These SDIPS generate an effective magnetic field in a diffusive S
in contact with a diffusive F, leading to the appearance of double
gap structures in the density of states (DOS) of both the S and F.
Within the terminology used above, this mechanism leads to the
anti-screening profile.

In general, in an S-F proximity system, the superconducting gap is
suppressed in the vicinity of the interface and recovers its
original value at a distance of a few $\xi_{S}$ from the
interface. \cite{Krivoruchko, Buzdin review} An induced magnetic
moment inside the S should further modify the DOS. Specifically,
the DOS is predicted to spin-split at a distance of a few
$\xi_{S}$ away from the interface. \cite{Cottet, Krivoruchko,
Halterman, Bergeret PRB72} In the anti-screening profile (and
assuming a positively spin-polarized F layer), the majority spins
will be shifted to lower energies and the minority spins will
shift to higher energies. The opposite process will occur in the
screening scenario, i.e., the majority spins will shift to higher
energies and the minority to lower energies. The magnitude of this
energy shift should be proportional to the induced moment or the
effective exchange field developed inside the S. Therefore,
split-gap structures or split bound states, measured in the DOS of
the S side near the interface, may provide evidence for an induced
magnetic moment. Moreover, the nature of this splitting holds
important information regarding the sign of the induced magnetic
moment inside the S. \cite{Bergeret PRB72}

All of the above theoretical works deal with conventional
\textit{s-wave} superconductors. To the best of our knowledge,
there are no theoretical predictions regarding the nature of this
effect in the high-$T_{c}$ cuprate superconductors. Specifically,
it is still unclear how the \textit{d-wave} symmetry, the short
coherence length and the comparable scale of the superconducting
gap and the exchange energy in such hetero-structures, will
influence the induced magnetization phenomena.

So far, only a few experimental works have identified possible
evidence for the induced magnetization effect. Stahn et. al.
\cite{Stahn} studied multilayers of $YBa_{2}Cu_{3}O_{7}$ and
$La_{2/3}Ca_{1/3}MnO_{3}$ using neutron reflectometry and provided
evidence for the screening profile. On the other hand, the
magnetization data of Stamopoulos et. al. \cite{Stamopoulos}
measured on Nb-manganites hybrids suggest anti-screening. A very
recent paper reports measurements of the polar Kerr effect in
Pb/Ni and Al/(Co-Pd) proximity systems and provides evidence for
the screening scenario. \cite{Palevski} Actually, one cannot argue
that these experiments contradict each other since one can not
compare all the system parameters that influence the sign of the
induced magnetic moment that were mentioned above. To the best of
our knowledge, the effect of induced magnetization on the DOS of
the S was not observed so far, and this is the main focus of this
paper.

\begin{figure}[t]
\includegraphics[width=8.4cm]{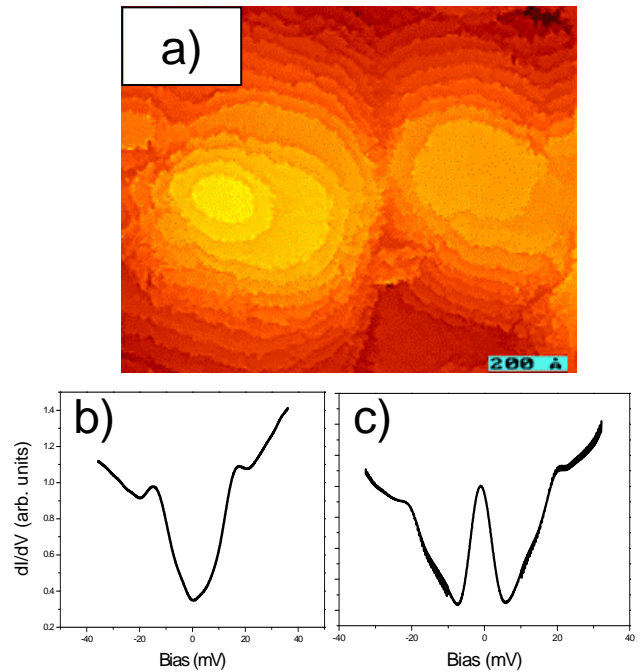}
\caption{(color online) (a) 100 $nm^{2}$ STM image of a bare
c-axis oriented YBCO film resolving unit-cell terraces. (b) and
(c) Typical tunneling spectra obtained on the bare c-axis films
and demonstrating the prevailing V-shaped gaps and the less
commonly observed ZBCP inside a gap. }\label{figure1}
\end{figure}

Previously, we performed tunneling spectroscopy measurements on
$SrRuO_{3}$ (SRO) ferromagnetic islands that were deposited on
nodal (110) $YBa_{2}Cu_{3}O_{7-\delta}$ (YBCO) surfaces.
\cite{Asulin} We found that the zero bias conductance peak (ZBCP),
the typical spectral feature measured on the nodal YBCO surface,
undergoes an anomalous splitting when measured on the F islands.
We attributed this effect to an induced magnetic moment in the S
layer. However, this is not yet a direct demonstration of the
effects of the induced magnetization on the DOS of S. In the
present work we focus on identifying signatures of induced
magnetic moment directly in the DOS of the S. We have performed
scanning tunneling spectroscopy of c-axis oriented YBCO films on
top of which SRO islands were deposited \textit{in-situ}. We have
detected an anomalous splitting of the superconducting gap feature
in the DOS of S when measured in the vicinity of an F island. The
maximal observed energy split, of about 5 meV, cannot be accounted
for by an effect of stray field emanating from the SRO
crystallites, but rather fits the scale of the SRO exchange
energy. Surprisingly, we found that the spatial extent of this
effect inside the S layer exceeds the YBCO coherence length by an
order of magnitude.

\section{Experiment}
Optimally-doped epitaxial c-axis oriented YBCO films were grown by
laser ablation deposition on (100) $SrTiO_{3}$ substrates. In
order to achieve a c-axis orientation of the films, the YBCO was
grown at a substrate temperature of $790^\circ$ C and a 100 mT
oxygen flow ambient. The films were cooled down in 0.5 atm of
oxygen pressure, with a dwell of 1 hour at $420^\circ$ C. Figure
1a presents a scanning tunneling microscope (STM) image of a bare
c-axis oriented film. The morphology of these films features
stacks of single c-axis unit-cell YBCO layers that form
terrace-shaped structures. These structures may emanate from a
screw dislocations growth mode. \cite{Wiesendanger} The bare
c-axis films showed a superconducting transition at $T_{c}\sim88$
K, implying slightly under-doped films, which is apparently due to
some oxygen loss from the surface. SRO islands were grown
epitaxially \textit{in-situ} on top of these c-axis surfaces and
under the same deposition conditions as for the bare YBCO films.
SRO is an itinerant ferromagnet with a Curie temperature of about
150 K for thin films. The similar lattice parameters of SRO and
YBCO \cite{Zakharov} facilitates epitaxial growth of
hetero-structures with a relatively high interface transparency.
Figure 2a presents a scanning electron microscope (SEM) image of a
0.5 nm thick (nominal) SRO layer over-coating a bare c-axis YBCO
film, illustrating the typical size and distribution of the SRO
islands (the bright rectangular crystallites). Their typical size
ranges between 50 and 100 nm and so is the spacing between them.
The thickness of the islands ranged between 2 and 4 SRO unit
cells. The STM image in Fig. 2b (obtained on the same sample)
reveals how the SRO islands are located on top of the bare c-axis
YBCO terraces, each contacting several terraces beneath it. This
is also illustrated in a 2-dimentional model of the system
sketched in Fig. 2c. The STM data presented here were all acquired
at 4.2 K, much lower than both the superconducting and
ferromagnetic transition temperatures, using a (normal metal)
Pt-Ir tip. The samples were transferred from the growth chamber in
a dry atmosphere and introduced into our cryogenic STM after being
exposed to ambient atmosphere for less than 2 minutes.

\begin{figure}[t]
\includegraphics[width=8cm]{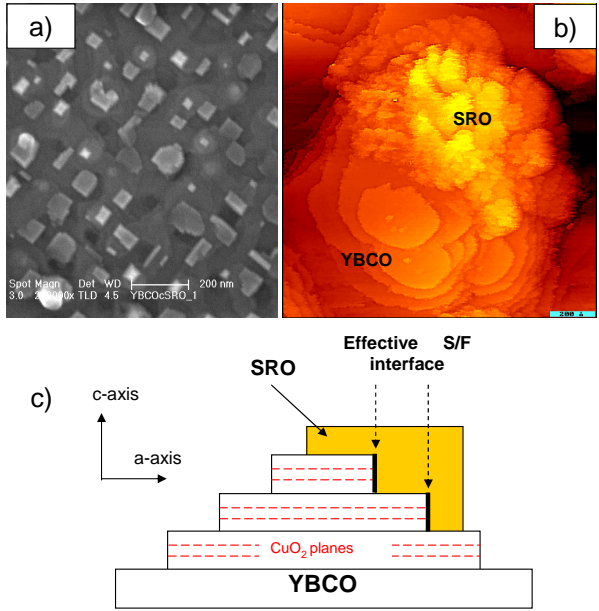}
\caption{(color online)(a) 0.9 $\mu m^{2}$ SEM image of a 0.5 nm
thick (nominal) SRO layer over-coating a bare c-axis film,
demonstrating the SRO-islands topography. (b) 120 $nm^{2}$ STM
image obtained on the same sample. (c) A schematic drawing of the
samples illustrating how the SRO island is situated on top of the
YBCO terraces. The bold black lines at the terrace edges beneath
the SRO island mark the effective S/F interfaces where the
$CuO_{2}$ planes are exposed.}\label{figure2}
\end{figure}

\section{RESULTS AND DISCUSSION}

\begin{figure}[t]
\includegraphics[width=8.5cm]{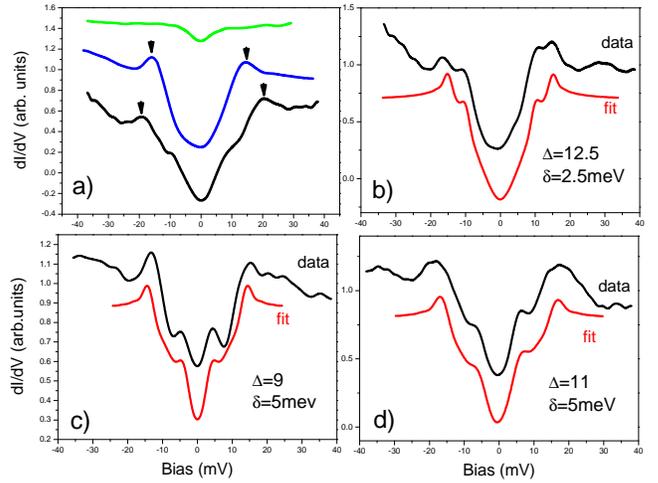}
\caption{(color online) Typical tunneling spectra obtained on a
c-axis YBCO film on top of which SRO islands were deposited
(nominally 0.5 nm thick SRO layer). (a) The upper green curve was
obtained on an SRO island, the middle blue curve was obtained at a
distance of about 30 nm away from the nearest SRO island while the
lower black curve was obtained in between. The upper (lower) curve
is shifted up (down) for clarity. The arrows mark the position of
the (outer, in the lower curve) coherence peaks. (b) to (d)
demonstrate the split-gap structures that were measured on the
YBCO surface in the vicinity of an SRO island (upper black
curves). The lower red curves (shifted down for clarity) are fits
to a phenomenological model as described in the text. The unsplit
gap values, $\Delta$, and the split amplitude, $\delta$, are
specified in each graph.}\label{figure3}
\end{figure}

The DOS of the bare c-axis YBCO films featured mainly V-shaped
gaps with a maximum value of $\Delta=17 meV$, as seen in Fig. 1b.
Less frequently, a ZBCP inside a gap structure was observed, as
depicted in Fig. 1c, reflecting a contribution of a nodal (110)
facet. These spectral features follow the known dependence of the
tunneling spectra on the local relative orientation of the
\textit{d-wave} OP and the normal to the surface. \cite{Kashiwaya,
Amos EPL} In a bilayer consisting of a high $T_{c}$ cuprate S and
a normal metal, the anisotropy of the \textit{d-wave} OP is
manifested also in the proximity effect. It is well established
that in such proximity systems, the penetration of the OP into the
normal metal is enabled primarily along the $CuO_{2}$ planes and
not along the c-axis direction [see Ref. \cite{Amos PRL} and
references therein]. Therefore, the proximity coupling between the
S and the F islands in our samples takes place mainly at the edges
of the YBCO terraces where the $CuO_{2}$ planes are exposed. These
edges constitute the most effective S/F interfaces in our system
and are marked by the bold black lines in the scheme of Fig. 2c.
The superconductor OP is expected to penetrate into the F islands
predominantly across these interfaces. Indeed, small gap-like
features with a very high zero-bias conductance were typically
measured in the DOS of the F islands as apparent in the upper
green curve of Fig. 3a. This is consistent with the conventional
picture of a short-range penetration of the OP into the F side in
S/F bilayers. \cite{Buzdin review} We have also measured the DOS
on the YBCO surface next to the SRO islands. This is equivalent to
measuring the S side of an S/F interface, where the most relevant
length-scale here is the distance from the above effective S/F
interfaces and the point of measurement. Far enough from the
interface, for example at a distance of 30 nm, V-shaped gaps
resembling those found on the bare YBCO films were measured
(middle curve in Fig. 3a).

\begin{figure}[t]
\includegraphics[width=6.5cm]{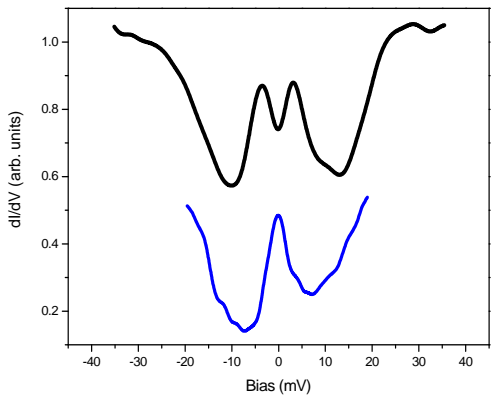}
\caption{(color online) The upper black curve demonstrates a
split-ZBCP obtained on the YBCO surface in the vicinity of an SRO
island. The lower blue curve was obtained on a (110) facet at a
distance of 25 nm away from the nearest SRO
island.}\label{figure4}
\end{figure}

Interestingly, however, the DOS in close vicinity to the SRO
islands featured a split-gap structure as depicted by the black
curves in Figs. 3a-d. The amplitude of the split, $\delta$, taken
as half the distance between the measured outer and inner peaks
(and also extracted from the fits discussed below), reaches a
maximum of 5 meV. Such a large split cannot be explained by an
effect of stray field emanating from the SRO crystallites. The
maximal stray field in the vicinity of an SRO crystallite cannot
exceed 0.2 Tesla. \cite{Klein APL1, Klein APL2} In order to
achieve energy shifts of a few meV, as observed here, a magnetic
field that is at least an order of magnitude larger is required.
On the other hand, the above energy scale of a few meV is of the
order of the exchange energy in SRO that is estimated to be
$E_{ex}\sim K_{B}T_{Curie}\sim 13 meV$. This value thus sets the
upper limit for the magnitude of the split at $\sim 6.5$ meV. We
would like to emphasize that the observed splitting is clearly an
effect related to the existence of the nearby F islands and is
absent in the bare c-axis films. Following the above
considerations, we attribute these split-gap structures to the
effect of an induced magnetic moment inside the S, consistent with
the above theoretical prediction of the splitting of the DOS.
\cite{Cottet, Krivoruchko, Halterman, Bergeret PRB72} The red
lower curves in Figs. 3b,c,d are fits to a phenomenological model
that takes into account the basic important physical ingredients
of the problem. Here, the total tunneling conductance is given by
\begin{equation}\label{eq:sigma}
    \sigma_{total}=P\cdot\sigma_{\uparrow}(E-\delta)+(1-P)\cdot\sigma_{\downarrow}(E+\delta),
    \end{equation}
where $\sigma_{\uparrow(\downarrow)}$ are given by the extended
BTK formalism \cite{Kashiwaya} for tunneling into a
\textit{d-wave} S, $\delta$ is an exchange-field induced splitting
parameter and P is a spin-polarization factor for the electrons
population. The latter parameter was found to be significantly
smaller compare to the values inside the SRO islands (see below).
Clearly, the main features of these split-gap structures are
nicely reproduced. The unsplit gap values, $\Delta$, (specified in
Figs. 3b,c,d) are smaller than the pristine gap value measured on
the bare c-axis films. This corresponds to the reduction of the OP
at the S side near the S/F interface mentioned above.
Nevertheless, for specific combinations of $\Delta$ and $\delta$,
the energy position of outer coherence peaks of the split gaps can
exceed that of the pristine gaps ($\Delta\sim17 meV$) as depicted
in the black curve in Fig. 3a. This point will be further
discussed below.

The effect of the induced magnetization also manifested in the
splitting of the less-abundantly observed ZBCP (black curve of
Fig. 4) measured in the vicinity of SRO islands. The maximum split
amplitude for the ZBCP was also about 5 meV. This value coincides
with the largest split amplitude that we observed in our previous
work on SRO islands deposited on (110) YBCO films. \cite{Asulin}
However, the two peaks of the split ZBCPs that were observed here
were almost of the same height, in contrast to the large asymmetry
observed in our previous work. \cite{Asulin} This is due to the
much larger polarization of the SRO islands compared to the small
polarization originating from the induced magnetization inside the
YBCO. Indeed, a small polarization factor of $0.5<P<0.55$ is
required in the fits we made for the measured split-gaps in Figs.
3b,c,d.

We note that based on our data, we cannot determine the sign of
the induced magnetic moment. To that end, a spin polarized STM is
required in order to detect whether the majority (or minority) of
spins were shifted to higher or lower energies. In addition,
further experiments are needed in order to resolve the open
question regarding the existence of triplet superconductivity near
the interface.

Surprisingly, the distance from the SRO islands over which the
split-gap and ZBCP features are observed, is an order of magnitude
larger than $\xi_{s}$ in YBCO, which is estimated to be 2-3 nm
along the $CuO_{2}$ planes. \cite{Segawa} Only at a distance of
about 30 nm away from any identifiable SRO island, the gap seems
to recover its full value and does not appear to be split (middle
blue curve in Fig. 3a taken at a distance of 30 nm away from the
nearest SRO island). The same holds for the ZBCP where the split
vanishes over a comparable length scale (the blue curve in Fig. 4
was obtained at a distance of 25 nm away from an SRO island). This
is inconsistent with the theoretically predicted penetration
length of the induced magnetic moment in S, of only a few
$\xi_{S}$ away from the S/F interface. One possible explanation
for this discrepancy is the existence of another relevant length
scale, that of the spin diffusion length. If a spin imbalanced
population of electrons is formed inside S in close vicinity to
the S/F interface, it is expected to loose its polarization at a
distance comparable to the spin diffusion length. In a
conventional S, this length scale is much shorter than the
coherence length (\textit{e.g.} a few 100 nm in Al). Therefore,
the physical size of the Cooper pairs, $\xi_{S}$, will determine
the effective penetration length of the induced magnetic moment
and spin diffusion will not play a role in this problem.
\cite{Bergeret PRB69} However, if the spin diffusion length is
larger than the coherence length, one would expect that the
imbalance in the spin population would be 'felt' by the S up to
distances that are larger than $\xi_{s}$. This is indeed the case
in YBCO for which the spin diffusion length was estimated to range
between 10 and 15 nm along the c-axis \cite{Soltan, Pena}, much
larger than $\xi_{s}$ in YBCO that is estimated to be 2-3 nm. One
can expect that along the $CuO_{2}$ planes the spin diffusion
length will be even larger and account for the penetration length
of the effect that is observed in our experiment.

\begin{figure}[t]
\includegraphics[width=8.5cm]{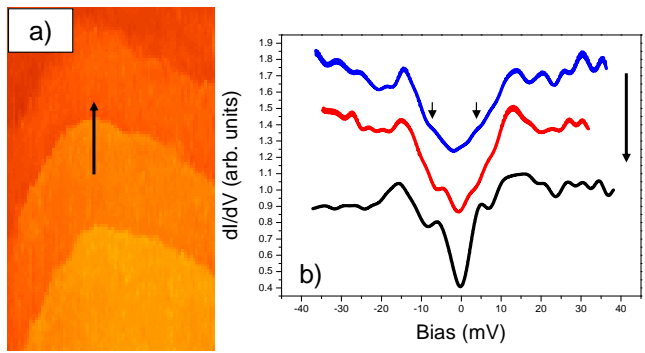}
\caption{(color online) (a) $30\times40 nm^{2}$ STM image of an
YBCO surface between SRO islands showing the YBCO terraces. (b)
Tunneling spectra obtained along the black arrow in (a).
}\label{figure5}
\end{figure}

The splits were most pronounced in the vicinity of the YBCO
terrace edges that expose the $CuO_{2}$ planes. Figure 5a presents
an STM image that zooms in on an YBCO surface in between SRO
islands and features these terraces. The spectra presented in Fig.
5b were acquired along the black arrow marked in Fig 5a. When the
DOS was measured on top of the terrace (upper curve in Fig. 5b),
only a weak signature of the split is observed inside the gap
(marked by arrows). When measured closer to the terrace edge, the
split becomes more pronounced, as evident from the lower two
curves. This is because in that case the STM tip coupling to the
$CuO_{2}$ planes, along which spin diffusion takes place, is more
effective. Moreover, the coupling to the nodal gapless excitations
is smaller on a terrace edge, thus the smearing of in-gap
structures (the inner peaks in our case) is reduced. \cite{Sharoni
PRB65, SharoniEPL62}

One can argue that the split gaps that we observed result from the
existence of a sub-dominant OP in our system. Indeed, double gap
structures were previously observed on Ca-doped YBCO surfaces and
were considered to originate from a sub-dominant OP. \cite{Sharoni
PRB65} However, this sub-dominant OP starts to emerge at the
optimal doping level of YBCO and is prominent only in the
over-doped regime \cite{Sharoni PRB65, Dagan} achieved effectively
by Ca-doping. Our bare c-axis samples were slightly under-doped
and do not show these double-gap features as mentioned above.
Moreover, a sub-dominant OP should only manifest itself as a
sub-gap structure and is not expected to shift the original
coherence peaks to larger values. In our measurements, in
contrast, the outer coherence peaks achieved values of up to 20
meV, which is higher than the maximal value observed on the bare
YBCO films (17 meV), as shown in Fig. 3a. In addition, if
interpreted by a sub-dominant OP, the position of the inner peaks
in our data correspond to an anomalously large sub-dominant OP
(more than 50 $\%$ of the YBCO gap), which is highly unlikely.

\section{summary}
In conclusion, our scanning tunneling spectroscopy measurements
show an anomalous splitting of the gap and ZBCP features when
measured on the YBCO high $T_{c}$ superconductor near the
interface with the itinerant ferromagnet SRO. The large split
magnitude, of up to 5 meV, can not be explained by an effect of
stray magnetic fields emanating from the SRO islands but is rather
in accord with the energy scale of the exchange field in SRO.
These findings are consistent with the recently predicted induced
magnetization effect in superconductor\ ferromagnet junctions and
constitute, to the best of our knowledge, the first observation of
its manifestation in the DOS of the superconductor.

\section{Acknowledgments}
\indent We thank G. Deutscher, D. Orgad and L. Klein for helpful
discussions. This work was supported by the Israel Science
Foundation, Center of Excellence program, (grant \# 481/07) and by
the German-Israeli Project Cooperation (DIP). O. M. acknowledges
the Harry deJur chair of applied sciences at the Hebrew University
and G. K. acknowledges the Karl Stoll chair in advanced materials
at the Technion.

\end{document}